\documentclass{iopart}
\usepackage{graphicx}
\usepackage{rotating}
\usepackage{longtable} 
\usepackage{SIunits}
\usepackage[usenames]{color}

\newcommand{\mosingle}{\ensuremath{\mathrm{La_{0.5}Sr_{1.5}MnO_4}}}
\newcommand{\lsmo}{\ensuremath{\mathrm{LaSr_{2}Mn_{2}O_7}}}
\newcommand{\oo}{\ensuremath{(\frac{1}{4},\frac{1}{4},0)}}
\newcommand{\af}{\ensuremath{(001)}}
\newcommand{\mo}{(0,0,1)~}
\renewcommand{\vec}[1]{\mbox{\boldmath$#1$}}

\begin{document} 
\letter{Separating the causes of orbital ordering in \lsmo ~using resonant soft x-ray diffraction}
\author{S.~B.~Wilkins$^{1}$, N.~Stoji\' c$^{2}$, T.~A.~W.~Beale$^{3}$, N.~Binggeli$^{2}$, P.~D.~Hatton$^{3}$, P.~Bencok$^{1}$, S.~Stanescu$^{1}$, J.~F.~Mitchell$^{4}$, P.~Abbamonte$^{5}$, M.~Altarelli$^{2,6}$} 
\address{$^{1}$European Synchrotron Radiation
  Facility, Bo\^\i te Postal 220, F-38043 Grenoble Cedex, France}
\address{$^{2}$Abdus Salam International Centre for Theoretical Physics, Trieste 34014, Italy}
\address{$^{3}$Department of Physics, University of Durham,
Rochester Building, South Road, Durham, DH1 3LE,
UK}
\address{$^{4}$Materials Science Division, Argonne National Laboratory, Argonne, Illinois 60439, USA}
\address{$^{5}$National Synchrotron Light Source, Brookhaven National Laboratory, Upton, NY 11973, USA}
\address{$^{6}$European XFEL Project Team, Desy, Notkerstra\ss e 85, 22607 Hamburg, Germany}

\ead{wilkins@esrf.fr}

\begin{abstract}
Resonant soft x-ray diffraction has been used to probe the temperature dependent orbital and magnetic structure of \lsmo. Previous crystallographic studies have shown that this material has almost no MnO$_{6}$ oxygen displacement due to Jahn-Teller distortions at low temperatures. 
Within the low-temperature A-type antiferromagnetic phase, we found strong intensity at
 the \oo\ orbital and \mo\ magnetic reflections. 
This shows that even in the near absence of Jahn-Teller distortion, this compound is strongly orbitally ordered.
A fit to the Mn $L$-edge  resonance spectra demonstrates the presence of orbital ordering of 
the Mn$^{3+}$ ions with virtually no Jahn-Teller crystal field in addition 
to  possible Mn$^{3+}$ and Mn$^{2+}$~like valence fluctuations. 
\end{abstract}



Ferromagnetism and charge ordering in the doped perovskite type manganese oxides R$_{1-x}$A$_{x}$MnO$_{3}$ (R = rare earth, A = Sr,Ca) have attracted considerable interest since the discovery of colossal magnetoresistance (CMR) \cite{jin:413}. Recently, much attention has been paid to the importance of the charge, lattice, spin and orbital degrees of freedom to explain the complex and anomalous structural, magnetic and transport behavior observed in the manganites. The interplay between these degrees of freedom can cause localisation of electrons on alternative manganese atoms to produce charge-ordered lattices. The ferromagnetism and metallic conductivity observed at low temperatures can be understood on the basis of the double-exchange mechanism whereby $e_{g}$ electrons hop between Mn sites through hybridisation with oxygen $2p$ orbitals and align the localised $t_{2g}$ spins by strong Hund's coupling \cite{zener:403,anderson:675,gennes:141}. However the understanding of the transport properties, such as CMR, and the complicated magnetic phase diagrams of the manganites requires a further ingredient, that of the orbital degree of freedom \cite{maezono:11583,mizokawa:R493}.

The study of the orbital ordering phenomenon is, therefore, vital for the understanding of the complex properties of the manganite systems.
Direct observation of orbital ordering in manganites has recently become possible using resonant soft x-ray diffraction at the Mn $L_{2,3}$ absorption edges after theoretical predictions by Castleton and Altarelli \cite{castleton:1033}.  Mn L$_{2,3}$ resonant x-ray scattering experiments have been performed in La$_{0.5}$Sr$_{1.5}$MnO$_{4}$ \cite{wilkins:167205,dhesi:056403} and in Pr$_{0.6}$Ca$_{0.4}$MnO$_{3}$ \cite{thomas:237204}, however in both cases the orbital degree of freedom is controlled by strong Jahn-Teller distortions.
In this letter we report experimental results on \lsmo. Previous crystallographic studies \cite{argyriou:15269,wilkins:205110}
have indicated that this system undergoes extremely small Jahn-Teller distortions at low temperature. 
Despite this our results  demonstrate the 
presence of long-range orbital order, in addition to magnetic order.
The fit to the orbital ordering spectrum using multiplet calculations in a crystal field shows 
the presence of a vanishing Jahn-Teller distortion in addition to possible Mn$^{3+}$ and Mn$^{2+}$~like valence fluctuations.

\begin{figure}
\begin{center}
\includegraphics[height=0.15\paperheight]{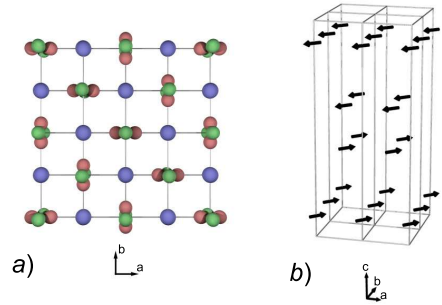}
\end{center}
\caption{The crystal structure of the bilayer manganite $\mathrm{La_{2-2x}Sr_{1+2x}Mn_{2}O_{7}}$ with $x=0.5$ at low temperature. (a) The arrangement of the previously proposed ``Mn$^{3+}$'' and ``Mn$^{4+}$'' manganese ions are shown within the tetragonal unit cell.  A plan view of the Mn$^{3+}$ orbitals within the $ab$ plane displaying orbital order of the $x^2-z^2,y^2-z^2$ type is given in (a). This is the dominant type of orbital ordering found in Ref.~\cite{koizumi:5589} for a doping x = 0.42. Our results suggest a checkerboard pattern closer to Mn$^{2+}$/Mn$^{3+}$ rather than Mn$^{3+}$/Mn$^{4+}$, however the alternating pattern remains the same. The arrangement of the magnetic spins of the Mn$^{3+}$ ions in the low temperature antiferromagnetic structure is shown schematically in (b).} 
\label{fig:cartoon}
\end{figure}

The samples were single crystals of the bilayer manganite \lsmo\ which were melt grown in flowing oxygen using a floating zone optical image furnace. The system La$_{2-2x}$Sr$_{1+2x}$Mn$_{2}$O$_{7}$ is a layered perovskite in which MnO$_{2}$ double layers and (La,Sr)$_{2}$O$_{2}$ blocking layers are stacked alternatively (see Fig~1 from Ref~\cite{wilkins:205110}). The reduced dimensionality causes the system to display a greatly enhanced magnetoresistance \cite{moritomo:141} and a reduced ferromagnetic transition temperature. Charge ordering has been reported in the temperature range from 100 to 200~K  existing only over a narrow compositional range $0.475<x<0.55$\cite{kimura:11081,li:R3205}. Below 170~K the $x=0.5$ material, \lsmo, adopts an A-type antiferromagnetic ordering of the Mn spins \cite{argyriou:15269} (see Fig.~\ref{fig:cartoon}) and crystallographic
 and high-energy x-ray diffraction studies have shown the disappearance\cite{argyriou:15269,kubota:1606} or reduction\cite{wilkins:187201} of the Jahn-Teller distortion, 
such that the MnO$_{6}$ octahedra are almost undistorted below $\sim$ 100~K.

Experiments were carried out using the in-vacuum diffractometer on beamline ID08 at the European Synchrotron Radiation Facility. A single crystal of \lsmo\ cut with the [110] direction normal to the sample surface was used for measurements of the \oo\ orbital order reflection. Calibration of the energy of the incident x-ray beam at ID08 by gas absorption gives an absolute accuracy of 0.2 eV at 640 eV. Measurements of the anti-ferromagnetic reflection \af, were performed on  a cleaved crystal with the [001] direction surface normal. The temperature dependancies of both these reflections were obtained using beamline X1B at the National Synchrotron Light Source. At both beamlines the experimental procedure was identical to our previous studies\cite{wilkins:187201,wilkins:167205}. 

\begin{figure}
\begin{center}
\includegraphics[height=0.15\paperheight]{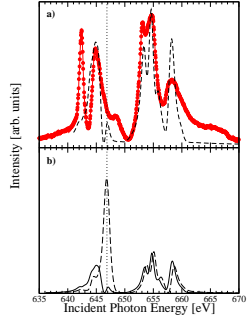}
\end{center}
\caption{(a) Scattered x-ray intensity as a function of incident photon energy at constant wavevector $\vec{Q}_{OO} = \oo$ (circles). The black dashed line shows the theoretical fit to the data. (b) Theoretical simulation (dashed black line) of the energy spectra with an 8-fold increase in the Jahn-Teller distortion. The fit to the experimental data is repeated for comparison (solid line).}
\label{fig:oo-resonance}
\end{figure}

\begin{figure}
\begin{center}
\includegraphics[height=0.15\paperheight]{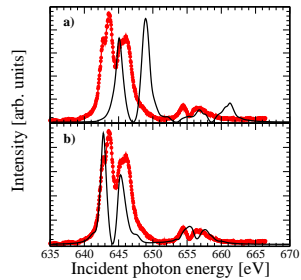}
\end{center}
\caption{ Scattered x-ray intensity as a function of incident photon energy at a constant wavevector $\vec{Q}_{AF} = \mo$ (red line with circles) with theoretical fits (solid black lines) describing the superposition of (a) Mn$^{3+}$/Mn$^{4+}$ and (b)  Mn$^{3+}$/Mn$^{2+}$ type.   }
\label{fig:af-resonance}
\end{figure}

At ID08, the sample was cooled to $\sim20$~K using liquid helium and an intense and narrow reflection was found at a wavevector of $\vec{Q}_{OO} = \oo$ at an incident energy of 643~eV. The solid circles in Fig.~\ref{fig:oo-resonance}a show the scattered intensity as a function of the incident photon energy at constant wavevector, through the manganese $L_{3}$ and $L_{2}$ edges. On first inspection, there appears to be two main features at the $L_{3}$ edge and three at the $L_{2}$ edge. In contrast to previous measurements on \mosingle\cite{wilkins:167205} and $\mathrm{Pr_{0.6}Ca_{0.4}MnO_{3}}$\cite{thomas:237204} the maximum scattered intensity is primarily observed at the $L_{2}$ edge rather than at the $L_{3}$ edge. Measurements of the A-type antiferromagnetic reflection in ${\rm La_{2-2x}Sr_{1+2x}Mn_2O_7}$ for $x=0.475$ at $\vec{Q}_{AF} = \mo$ have been reported by Wilkins {\it et al.}\cite{wilkins:187201} at a temperature of 83~K. In Fig.~\ref{fig:af-resonance}(a,b) we show the scattered intensity measured at 20~K in \lsmo\  as a function of energy at constant wavevector. In this case, at the $L_{2}$ edge little scattering was observed but at the $L_{3}$ edge very strong intensity was found which is comprised of two main features. Our experimental data in Figs.~\ref{fig:oo-resonance} and \ref{fig:af-resonance} indicate that \lsmo~ is simultaneously orbitally and magnetically ordered at low temperature. Soft X-ray diffraction is not particularly surface-sensitive\cite{wilkins:187201}. The inverse width in reciprocal space of the reflections gives an indication of the penetration depth. This indicates we are probing thousands of $\mathrm{\AA}$ngstroms into the crystal. 

In order to identify the origin of the resonant x-ray scattering signal, we have performed multiplet 
calculations in a crystal field. On the Mn$^{3+}$ site with  $D_{4h}$ symmetry, two crystal field parameters
are to be acquired from the fitting procedure: cubic ($X^{400}$) and tetragonal ($X^{220}$). 
In the absence of the experimental evidence, we assume that the spins are  aligned
in the [110] direction, which lowers the symmetry to that of the $C_i$ point group. 
Choosing this direction as quantization axis, the resonant scattering amplitude at the orbital ordering 
wave vector \oo, is proportional to the following combination
of the atomic scattering tensors \cite{StoBinAlt05}:

\begin{equation}
\label{f_OO}
f^{\rm OO}_{\rm res} \propto F^{\rm e}_{0;1}- 
F^{\rm e}_{0;-1}+F^{\rm e}_{1;0}-F^{\rm e}_{-1;0},
\end{equation}
with $ F^{\rm e}_{m;m'}$  defined as:
\begin{equation}
F^{\rm e}_{m;m'}=\sum_n \frac{\langle 0|J^{1\dag}_{m}|n\rangle \langle n|J^{1}_{m'}|0\rangle}{E_0-E_n+\hbar \omega + i\Gamma /2},
\end{equation}
where $m$ and $m'$ denote polarization states and $J^1_m$ are the dipole operators defined in spherical coordinates. $|0\rangle$ represents the ground state with energy $E_0$ and
$|n\rangle$ intermediate states with energy $E_n$. The photon energy is $\hbar \omega$  and $\Gamma$
stands for the broadening due to the core-hole lifetime. 
Similarly, for the magnetic scattering with the wave vector \mo, the scattering amplitude can be expressed as:
\begin{equation}
\label{f_MO}
f_{\rm res}^{\rm MO} \propto 
F^{\rm e}_{1;1}-F^{\rm e}_{-1;-1}.
\end{equation}

The Slater integrals of the $d$-$d$ and $p$-$d$ direct and exchange interactions
were scaled down to 75\% of their atomic value. The $p$-shell spin-orbit parameter
was increased by 9\% from the Hartree-Fock value to correspond to the experimental
value \cite{ThoAttGul01}. 
We used $\Gamma=$0.5~eV for the Lorentzian broadening due to the core-hole lifetime. 
In addition, to take into account also the experimental
energy resolution,  the scattering intensity, 
$I(\hbar \omega) \propto | f_{\rm res} |^2$, was convoluted with a Gaussian of width 0.1~eV. 

The best fit to the orbital ordering is shown in Fig. \ref{fig:oo-resonance}a. The corresponding
crystal field parameters are 
$ \rm X^{400}$ = 3~eV and 
$ \rm X^{220}$ = 0.4~eV, (or 10$\rm D_q$ = 0.91~eV and $\rm D_s$ = $-$0.048~eV).
The fit did not significantly change for variations of the cubic field $ \rm X^{400}$ in the interval 3$-$4~eV
and for the tetragonal (Jahn-Teller) field  $ \rm X^{220}$ in the interval 0.1$-$0.6~eV.
The fit displays a fair general agreement with the experimental spectrum. The obtained $L_3/L_2$ ratio
is satisfactory and
we reproduce most of the structure from the experimental spectrum. 
However, the first peak in the $L_3$ edge is not reproduced \cite{note2}
and broadened high-energy
features at both edges are missing in the fit.
 These wide shoulders may be related to band-structure effects, which are not incorporated in our model. In Fig. \ref{fig:oo-resonance}b
we illustrate the effect of an 8-fold increase of the Jahn-Teller field.
As shown before, \cite{wilkins:245102}, the $L_3$/$L_2$ ratio is becoming larger with the increase in the tetragonal component 
of the crystal field. In our calculations a small Jahn-Teller tetragonal crystal field is necessary to lift the degeneracy 
of the $e_{g}$ levels, however the tetragonal field included is extremely small.
 Its value is one order of magnitude smaller than that obtained, {\it e.g.}, for \mosingle\ \cite{wilkins:245102}, so that,
except for the lifting of orbital degeneracy, the Mn$^{3+}$ ion is essentially in a cubic field.
 This  shows we are within a regime where the scattering is dominated by 
orbital ordering of the $e_{g}$ electrons. 
 
We were not able to obtain a good fit to the magnetic scattering with  Mn$^{3+}$ ions alone, as it gave a single
peak at each edge. 
Since the $\vec{Q}_{AF} = \mo$ sees a superposition of \emph{all} manganese ions within the $ab$ plane, we 
included an additional type of Mn ion in our model. Figure~\ref{fig:af-resonance}a shows the  magnetic spectrum 
obtained by superposing the contributions of Mn$^{3+}$ and Mn$^{4+}$ with equal weights. The Mn$^{4+}$ ion induces a peak at
higher energy (at both edges) relative to the Mn$^{3+}$ ion,
as expected for ions
with larger oxidation number\cite{SchHenCre95}. This gives rise to a spectrum with two peaks at each edge, but
shifted to a higher energy relative to the experimental spectrum. Moreover, the ratio between the two main features at both the $L_{3}$ and $L_{2}$ edges is inverted. In order to reproduce the feature at low energy,
with a ratio consistent with experiment, we need to consider a Mn ion with lower oxidation number.
 In panel b) of Fig.~\ref{fig:af-resonance} we show the spectrum obtained with a 1:1 ratio of
Mn$^{2+}$ and Mn$^{3+}$. In these calculations, the Mn$^{3+}$ crystal field parameters were taken to be exactly the same as  
for the orbital spectrum in Fig.~\ref{fig:oo-resonance}a, and the  Mn$^{2+}$ and Mn$^{4+}$ spectra
were calculated in a purely cubic crystal field ($\mathrm{X^{400}} = 3.0~\mathrm{eV})$. 
In Fig.~\ref{fig:af-resonance}, we positioned the Mn$^{2+}$ (Mn$^{4+}$) edge, relative to the Mn$^{3+}$ edge, using the
theoretical chemical shifts obtained from our atomic multiplet calculations, {\it i.e.} $\sim$ -2~eV and 4~eV, respectively. 
Experimental L-edge absorption results on Mn$^{2+}$, Mn$^{3+}$, and Mn$^{4+}$ reference compounds show comparable chemical
shifts (between -1 and -2~eV for Mn$^{2+}$ and between 2 and 3~eV for Mn$^{4+}$ \cite{CraGroMa91,MorGroGla04,KobUsuIku04}). 

Surprisingly, much better agreement with the experimental data in Fig.~\ref{fig:af-resonance}(a,b) is obtained by the 
calculation for the Mn$^{2+}$/Mn$^{3+}$ case than for the Mn$^{3+}$/Mn$^{4+}$ case.  This may simply reflect the basic limitations of the atomic 
multiplet calculations (which, for one thing, only allow integer values for the valence) to describe an extended system 
with non-negligible interatomic hopping; on the other hand, it is noteworthy that an average valence between 2+ and 3+ 
has recently been suggested by Hartree-Fock calculations \cite{FerTowLit03,ZhePat03}, and experimental data which indicate 
that the additional holes reside more on the oxygen ligands \cite{JuSohKri97,SubGarSan02}. Such a change in the manganese valency from 3+/4+ to 2+/3+ would not alter the orbital occupancy of the  Mn$^{3+}$ ion. Both Mn$^{2+}$ and Mn$^{4+}$ are not Jahn-Teller active.

\begin{figure}\begin{center}
\includegraphics[height=0.15\paperheight]{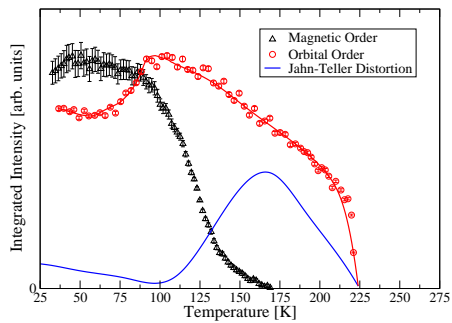}
\end{center}
\caption{Temperature dependence of the integrated intensity of the \oo\ orbital order reflection and the \af\ magnetic order reflection. The temperature depedence of the {\ensuremath{(\frac{3}{4},-\frac{1}{4},5)}} Jahn-Teller distortion peak, a measure of the amplitude of the Jahn-Teller distortion, is taken from Ref~\cite{wilkins:205110}\label{fig:temp}.}
\end{figure}

Figure 4 shows the temperature dependence of the orbital and magnetic reflections. 
The \oo\ reflection is observed below $T_{OO}$ which is coincident with $T_{JT}$. The Jahn-Teller distortions maximise at $T_N$ ($\sim$170~K) and decrease below this temperature due to the occurance of antiferromagnetic interactions. Note that the Jahn-Teller distortions are very weak below 100~K where the orbital and magnetic reflections are very strong. We note that below 100~K the Jahn-Teller distortions increase slightly, which coincides with a change in gradient of the intensity of the orbital and magnetic reflections.

In conclusion, we have reported the results of  resonant soft x-ray scattering studies of the  
orbital and antiferromagnetic ordering in \lsmo. 
The data of Fig.~\ref{fig:oo-resonance} and \ref{fig:af-resonance} immediately show that \lsmo\ 
is simultaneously both orbitally and antiferromagnetically ordered at low temperatures, despite the fact that
the Jahn-Teller distortions in this system are measured to be very small. This is further supported by 
 the theoretical fits, which display a good agreement with the experimental data for very small values of 
tetragonal crystal field. From these we can conclude that it is possible to obtain a strongly orbitally ordered phase within an A-type antiferromagnetic configuration in the absence of significant Jahn-Teller distortions. In such case, the energetics of the orbitally ordered 
configuration is favored by the magnetic interactions, originating from the superexchange mechanism, as described by
Goodenough, \cite{Goodenough:1963}.  The fit to the magnetic scattering data
suggests a  fluctuating valence situation, with Mn ions with valence charge between +3 and +2.

This work was supported by the Synchrotron Radiation Related Theory Network, SRRTN, of the EU.  N.S. gratefully acknowledges  the assistance of Paolo Carra in learning how to use the Cowan and ``Racah'' codes.  We thank F. M. F. de Groot for  helpful discussions. 
We are grateful for support from EPSRC for a studentship for T.A.W.B. and for a travel grant to NSLS.

\section*{References}
\bibliography{mang}

\begin{thebibliography}{10}

\bibitem{jin:413}
S.~Jin, T.~H. Tiefel, M.~McCormack, R.~A. Fastnacht, R.~Ramesh, and L.~H. Chen.
\newblock {\em Science}, 264:413, 1994.

\bibitem{zener:403}
C~Zener.
\newblock {\em Physical Review}, 82:403, 1951.

\bibitem{anderson:675}
P.~W. Anderson and H.~Hasegawa.
\newblock {\em Physical Review}, 100:675, 1955.

\bibitem{gennes:141}
P.~G. de~Gennes.
\newblock {\em Physical Review}, 118:141, 1960.

\bibitem{maezono:11583}
Ryo Maezono, Sumio Ishihara, and Naoto Nagaosa.
\newblock {\em Physical Review B (Condensed Matter and Materials Physics)},
  58:11583--11596, 1998.

\bibitem{mizokawa:R493}
T.~Mizokawa and A.~Fujimori.
\newblock {\em Physical Review B (Condensed Matter)}, 56:R493--R496, 1997.

\bibitem{castleton:1033}
C.~W.~M. Castleton and M.~Altarelli.
\newblock {\em Physical Review B (Condensed Matter and Materials Physics)},
  62:1033--1038, 2000.

\bibitem{wilkins:167205}
S.~B. Wilkins, P.~D. Spencer, P.~D. Hatton, S.~P. Collins, M.~D. Roper,
  D.~Prabhakaran, and A.~T. Boothroyd.
\newblock {\em Physical Review Letters}, 91:167205, 2003.

\bibitem{dhesi:056403}
S.~S. Dhesi, A.~Mirone, C.~De Nadai, P.~Ohresser, P.~Bencok, N.~B. Brookes,
  P.~Reutler, A.~Revcolevschi, A.~Tagliaferri, O.~Toulemonde, and G.~van~der
  Laan.
\newblock {\em Physical Review Letters}, 92:056403, 2004.

\bibitem{thomas:237204}
K.~J. Thomas, J.~P. Hill, S.~Grenier, Y-J. Kim, P.~Abbamonte, L.~Venema,
  A.~Rusydi, Y.~Tomioka, Y.~Tokura, D.~F. McMorrow, G.~Sawatzky, and M.~van
  Veenendaal.
\newblock {\em Physical Review Letters}, 92:237204, 2004.

\bibitem{argyriou:15269}
D.~N. Argyriou, H.~N. Bordallo, B.~J. Campbell, A.~K. Cheetham, D.~E. Cox,
  J.~S. Gardner, K.~Hanif, A.~dos Santos, and G.~F. Strouse.
\newblock {\em Physical Review B (Condensed Matter and Materials Physics)},
  61:15269--15276, 2000.

\bibitem{wilkins:205110}
S.~B. Wilkins, P.~D. Spencer, T.~A.~W. Beale, P.~D. Hatton, M.~v.~Zimmermann,
  S.~D. Brown, D.~Prabhakaran, and A.~T. Boothroyd.
\newblock {\em Physical Review B (Condensed Matter and Materials Physics)},
  67:205110, 2003.

\bibitem{moritomo:141}
Y.~Moritomo, A.~Asamitsu, H.~Kuwahara, and Y.~Tokura.
\newblock {\em Nature}, 380:141, 1996.

\bibitem{kimura:11081}
T.~Kimura, R.~Kumai, Y.~Tokura, J.~Q. Li, and Y.~Matsui.
\newblock {\em Physical Review B (Condensed Matter and Materials Physics)},
  58:11081--11084, 1998.

\bibitem{li:R3205}
J.~Q. Li, Y.~Matsui, T.~Kimura, and Y.~Tokura.
\newblock {\em Physical Review B (Condensed Matter and Materials Physics)},
  57:R3205--R3208, 1998.

\bibitem{kubota:1606}
M.~Kubota, H.~Fujioka, K.~Hirota, K.~Ohoyama, Y.~Moritomo, H.~Yoshizawa, and
  Y.~Endoh.
\newblock {\em J. Phys. Soc. Jpn.}, 69:1606, 2000.

\bibitem{wilkins:187201}
S.~B. Wilkins, P.~D. Hatton, M.~D. Roper, D.~Prabhakaran, and A.~T. Boothroyd.
\newblock {\em Physical Review Letters}, 90:187201, 2003.

\bibitem{koizumi:5589}
Akihisa Koizumi, Satoru Miyaki, Yukinobu Kakutani, Hiroyasu Koizumi, Nozomu
  Hiraoka, Kenji Makoshi, Nobuhiko Sakai, Kazuma Hirota, and Yoichi Murakami.
\newblock {\em Physical Review Letters}, 86:5589--5592, 2001.

\bibitem{StoBinAlt05}
N.~Stoji\'c, N.~Binggeli, and M.~Altarelli.
\newblock {\em Physical Review B (Condensed Matter and Materials Physics)},
  72:104108, 2005.

\bibitem{ThoAttGul01}
A.~Thomson and et~al.
\newblock {\em X-Ray Data Booklet}.
\newblock Lawrence Berkeley National Laboratory, University of California,
  Berkeley, CA 94720, January 2001.

\bibitem{note2}
We were able to reproduce the first peak in the $L_3$ edge by increasing the
  cubic crystal field parameter. But, simultaneously, it was necessary to
  additionally significantly increase the 2$p$ spin-orbit parameter. As such
  increase could not be justified and is in contrast with results from Ref.
  \cite{wilkins:245102}, we do not present the corresponding fits. It should be
  noted, however, that even in this case the conclusion concerning the
  Jahn-Teller field remains: only a very small tetragonal field could reproduce
  the observed spectrum.

\bibitem{wilkins:245102}
S.~B. Wilkins, N.~Stoji\' c, T.~A.~W. Beale, N.~Binggeli, C.~W.~M. Castleton,
  P.~Bencok, D.~Prabhakaran, A.~T. Boothroyd, P.~D. Hatton, and M.~Altarelli.
\newblock {\em Physical Review B (Condensed Matter and Materials Physics)},
  71:245102, 2005.

\bibitem{SchHenCre95}
P.~F. Schofield, C.~M. Henderson, G.~Cressey, and G.~van~der Laan.
\newblock {\em J. Synchrotron Rad.}, 2:93, 1995.

\bibitem{CraGroMa91}
S.~P. Cramer, F.M.~F. de~Groot, Y.~Ma, C.~T. Chen, F.~Sette, C.~A Kipke, D.~M.
  Eichhorn, M.~K. Chan, W.~H. Armstrong, E.~Libby, G.~Christou, S.~Booker,
  V.~McKee, O.~C. Mullins, and J.~C. Fuggle.
\newblock {\em J. Am. Chem. Soc.}, 113:7937, 1991.

\bibitem{MorGroGla04}
F.~Morales, F.M.~F. de~Groot, P.~Glatzel, E.~Kleimenov, H.~Bluhm, M.~Havecker,
  A.~Knop-Gericke, and B.~M. Weckhuysen.
\newblock {\em J. Phys. Chem. B}, 108:16201, 2004.

\bibitem{KobUsuIku04}
S.~Kobayashi, T.~Usui, H.~Ikuta, Y.~Uchimoto, and M.~Wakihara.
\newblock {\em J. Mater. Res.}, 19:2421, 2004.

\bibitem{FerTowLit03}
V.~Ferrari, M.~Towler, and P.~B. Littlewood.
\newblock {\em Physical Review Letters}, 91:227202, 2003.

\bibitem{ZhePat03}
G.~Zheng and C.~H. Patterson.
\newblock {\em Physical Review B}, 67:220404, 2003.

\bibitem{JuSohKri97}
H.~L. Ju, H.-C. Sohn, and K.~M. Krishnan.
\newblock {\em Physical Review Letters}, 79:3230, 1997.

\bibitem{SubGarSan02}
G.~Subias, J.~Garcia, M.~C. Sanchez, J.~Blasco, and M.~G. Proietti.
\newblock {\em Surface Review and Letters}, 9:1071, 2002.

\bibitem{Goodenough:1963}
J.~B. Goodenough.
\newblock {\em Magnetism and the Chemical Bond}.
\newblock Interscience, New York, 1963.

\end{thebibliography}
\end{document}